\documentclass[12pt,preprint]{aastex}

\newcommand{\Msun}{\mbox{\,M$_\odot$}}

\newcommand{\mic}{\mbox{$\,\mu$m}}

\newcommand{\fion}[2]{[{#1}\,{\sc {#2}}]}
\newcommand{\ltsimeq}{\raisebox{-0.6ex}{$\,\stackrel
        {\raisebox{-.2ex}{$\textstyle <$}}{\sim}\,$}} 
\newcommand{\gtsimeq}{\raisebox{-0.6ex}{$\,\stackrel
	{\raisebox{-.2ex}{$\textstyle >$}}{\sim}\,$}}

\newcommand{\rs}{RS~Oph}

\slugcomment{Submitted to ApJL}

\shorttitle{Spitzer observations of RS~Oph}
\shortauthors{Evans et al.}

\begin{document}

%% LaTeX will automatically break titles if they run longer than
%% one line. However, you may use \\ to force a line break if
%% you desire.

\title{Spitzer and ground-based infrared observations of the 2006 eruption of
RS~Ophiuchi}

%% Use \author, \affil, and the \and command to format
%% author and affiliation information.
%% Note that \email has replaced the old \authoremail command
%% from AASTeX v4.0. You can use \email to mark an email address
%% anywhere in the paper, not just in the front matter.
%% As in the title, use \\ to force line breaks.

\author{A. Evans\altaffilmark{1},
C. E. Woodward\altaffilmark{2},
L. A. Helton\altaffilmark{2},
R. D. Gehrz\altaffilmark{2},
D. K. Lynch\altaffilmark{3},
R. J. Rudy\altaffilmark{3},
R. W. Russell\altaffilmark{3},
T. Kerr\altaffilmark{4},
M. F. Bode\altaffilmark{5},
M. J. Darnley\altaffilmark{5},
S. P. S. Eyres\altaffilmark{6},
T. R. Geballe\altaffilmark{7},
T. J. O'Brien\altaffilmark{8},
R. J. Davis\altaffilmark{8},
S. Starrfield\altaffilmark{9},
J.-U. Ness\altaffilmark{9},
J. Drake\altaffilmark{10}
J. P. Osborne\altaffilmark{11},
K. L. Page\altaffilmark{11}
G. Schwarz\altaffilmark{12}
J. Krautter\altaffilmark{13}
}

\altaffiltext{1}{Astrophysics Group, Keele University, Keele, Staffordshire, ST5 5BG, UK,
{\it ae@astro.keele.ac.uk}}

\altaffiltext{2}{Department of Astronomy, School of Physics \& Astronomy, 116 Church
      Street S.E., University of Minnesota, Minneapolis, MN 55455, USA,
{\it chelsea@astro.umn.edu, ahelton@astro.umn.edu, gehrz@astro.umn.edu}}

\altaffiltext{3}{The Aerospace Corporation, Mail Stop M2-266, P.O. Box 92957, Los
      Angeles, CA~90009-2957, USA,
{\it David.K.Lynch@aero.org, Richard.J.Rudy@aero.org, Ray.Russell@aero.org}}

\altaffiltext{4}{Joint Astronomy Centre, 660 N. A'ohoku Place, University Park,
      Hilo, Hawaii 96720, USA,
{\it t.kerr@jach.hawaii.edu}}

\altaffiltext{5}{Astrophysics Research Institute, Liverpool John Moores University,
      Twelve Quays House, Birkenhead CH41 1LD, UK,
{\it mfb@astro.livjm.ac.uk, mjd@astro.livjm.ac.uk}}

\altaffiltext{6}{Centre for Astrophysics, University of Central Lancashire,
      Preston, PR1 2HE, UK, {\it spseyres@uclan.ac.uk}}

\altaffiltext{7}{Gemini Observatory, 670 N. A'ohoku Place, Hilo, HI\,96720, USA,
{\it tgeballe@gemini.edu}}

\altaffiltext{8}{Department of Physics \& Astronomy, University of Manchester,
      Manchester, UK, {\it tob@jb.man.ac.uk, rjd@jb.man.ac.uk}}

\altaffiltext{9}{School of Earth \& Space Exploration, Arizona State University,
      PO Box 871404, AZ 85287-1404, USA,
 {\it sumner.starrfield@asu.edu, ness@susie.la.asu.edu}}

\altaffiltext{10}{Harvard-Smithsonian Center for Astrophysics (CfA), 60 Garden
      Street, Cambridge, MA 02138, USA. {\it jdrake@cfa.harvard.edu}}

\altaffiltext{11}{Department of Physics and Astronomy, University of Leicester,
      Leicester, LE1 7RH, UK, {\it julo@star.le.ac.uk, kpa@star.le.ac.uk}}

\altaffiltext{12}{Steward Observatory, University of Arizona, 933 North Cherry
      Avenue, Tucson, AZ 85721, USA, {\it gschwarz@as.arizona.edu}}

\altaffiltext{13}{Landessternwarte, K\"{o}nigstuhl, D-69117 Heidelberg, Germany,
     {\it j.krautter@lsw.uni-heidelberg.de}}

% \author{}
% \affil{}
% \email{}

%% Notice that each of these authors has alternate affiliations, which
%% are identified by the \altaffilmark after each name.  Specify alternate
%% affiliation information with \altaffiltext, with one command per each
%% affiliation.

% \altaffiltext{1}{Visiting Astronomer, Cerro Tololo Inter-American Observatory.
% CTIO is operated by AURA, Inc.\ under contract to the National Science
% Foundation.}

%% Mark off your abstract in the ``abstract'' environment. In the manuscript
%% style, abstract will output a Received/Accepted line after the
%% title and affiliation information. No date will appear since the author
%% does not have this information. The dates will be filled in by the
%% editorial office after submission.

\begin{abstract}
We present Spitzer Space Telescope and complementary ground-based infrared
observations of the recurrent nova {\rs}iuchi, obtained
over the period 64-111~days after the 2006 eruption. The Spitzer IRS data show
a rich emission line spectrum superimposed on a free-free continuum. The
presence of fine structure and coronal infrared lines lead us to deduce that
there are at least two temperatures ($1.5\times10^5$~K and $9\times10^5$~K) in
the ejecta/wind environment, and that the electron density in the `cooler'
region is $2.2\times10^5$~cm$^{-3}$. The determination of elemental abundances
is not straightforward but on the assumption that the Ne and O fine structure
lines arise in the same volume of the ejecta, the O/Ne ratio is $\gtsimeq0.6$
by number.
\end{abstract}


% \object{RS~Ophiuchi}

\section{Introduction}

\rs\ is a recurrent nova (RN) that erupted in
1898, 1933, 1958, 1967, 1985, and possibly 1907 and 1945. The system
consists of a semi-detached binary system comprised of a
roche-lobe-filling giant (RG) and a massive ($\sim1.2\Msun$) white dwarf
\citep[WD;][]{shore,fekel}. 
As in classical novae, the eruption follows a thermonuclear
runaway on the surface of the WD \citep{tnr}. In the case
of the \rs\ class of RNe, however, the ejected material runs into, and shocks,
a dense RG wind \citep{bk85}.

The 1985 eruption of \rs\ was, for the first time, the subject of a
multi-wavelength observational campaign, from the radio to the x-ray
\citep{vnu}. The most recent eruption, on 2006 February 12.83
\citep[][we take this to define the origin of time: $t=0$]{hirosawa},
triggered an even more intensive campaign. Infrared (IR)
observations by \citet{das06} and \citet[][hereafter Paper~I]{evans07} 
show evidence for the shock, seen also at radio \citep{obrien}
and x-ray \citep{bode06,sokoloski,ness,osborne} wavelengths, as the
ejecta interact with the RG wind.
HST observations of the ejecta from the
2006 eruption \citep{bode07} showed that emission in \fion{O}{iii}5007\AA\
consists of two `rings' consistent with constant velocity expansion
of the radio lobes seen earlier in the eruption \citep{obrien}. In contrast
the \fion{Ne}{v}3426\AA\ lines possibly arise from two `polar caps' at the
extremities of the rings.

We present here further IR observations, obtained with the 
Infrared Spectrometer \citep[IRS;][]{houck} on the Spitzer Space Telescope
\citep{werner, gehrz07} and from ground-based facilities.

\section{Observations} 

\subsection{Spitzer}

\rs\ was observed on 2006 Apr 16 ($t=62.51$~days) and 2006 Apr 26
($t=72.54$~days) with {\it Spitzer} IRS as part of the Directors Discretionary
Time program PID~270. Observations were performed using all IRS modules, with
IRS blue (13.3--18.6\mic) peak-up on \rs. The spectroscopy consisted of 5 cycles
of 14 second ramps in short-low mode, 5 cycles of 30 second ramps in both
short-high and long-low modes, and 5 cycles of 60 second ramps in long-high
mode. IRS basic calibrated data products (BCDs) were processed with version
14.0.0 of the IRS pipeline. Details of the
calibration and raw data processing are specified in the IRS Pipeline
Description Document\footnotemark, version
1.0.\footnotetext{ssc.spitzer.caltech.edu/irs/dh/PDD.pdf} 

\begin{figure}
\includegraphics[angle=-90,scale=.50]{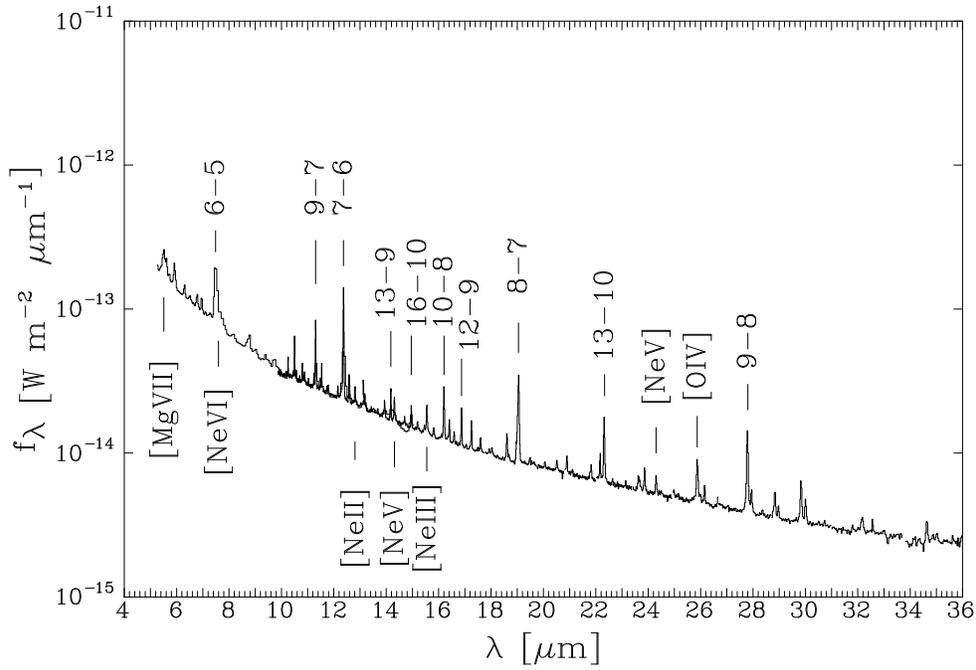}
\caption{Spitzer spectrum of \rs\ for 2006 April 16 UT. Principal H
recombination ($n-m$) and fine structure lines are identified;
many of the lines without identification are also H recombination
lines originating in high ($n\ge15$) levels.}
\label{spitzer}
\end{figure}

Bad pixels were interpolated in individual BCDs using bad pixel masks
provided by the Spitzer Science Center (SSC). For the low resolution 
data, multiple data collection events were obtained at two different positions on the
slit using {\it Spitzer's} nod function. The low resolution two-dimensional BCDs were
differenced to remove the background flux contribution and the data were extracted
with SPICE \citep[version 1.3-beta1;][]{spice} using the default point source
extraction widths. For the high resolution data, the point spread function
nearly fills the high resolution slit length, so it is not possible to perform background
subtraction using nod pairs. Instead, separate sky observations were used to
construct a master sky that was subtracted from the individual BCDs to
remove the background contribution. Extraction was performed in the same manner
as for the low resolution modules. For both resolution regimes the extracted, 
background-corrected data were combined, using a weighted linear mean, into a
single output data file. The point-to-point errors were estimated from the
standard deviation of the flux at each wavelength bin except when there were
fewer than three data points in which case the errors generated by the SSC
pipeline were added in quadrature to determine the final error. For the high resolution
modules, in the order overlap regions, the long wavelength
edge of the orders were much less reliable than the short wavelength edges, and
so only the short wavelength regions of overlap were retained, i.e. the data
from the lower order. As suggested by the SSC, data outside the ranges
5.2--14.5\mic\ for the SL module, 9.9--19.6\mic\ for the SH module, and
14.0--38.0\mic\ and 18.7--37.0\mic\ for the LL and LH modules, respectively, were
discarded as unreliable. The spectral lines were fit using a least squares
Gaussian routine that fit the line center, line amplitude, continuum amplitude
and the slope of the continuum.
Individual segments of the IRS spectrum were shifted upwards or downwards by
$\ltsimeq10$\% to ensure that segments adjoined smoothly. 
The spectrum for 2006 April 16 is shown in Fig.~\ref{spitzer}; the spectrum for
2006 April 26 is not substantially different.

\subsection{Ground-based observations}

\begin{figure}
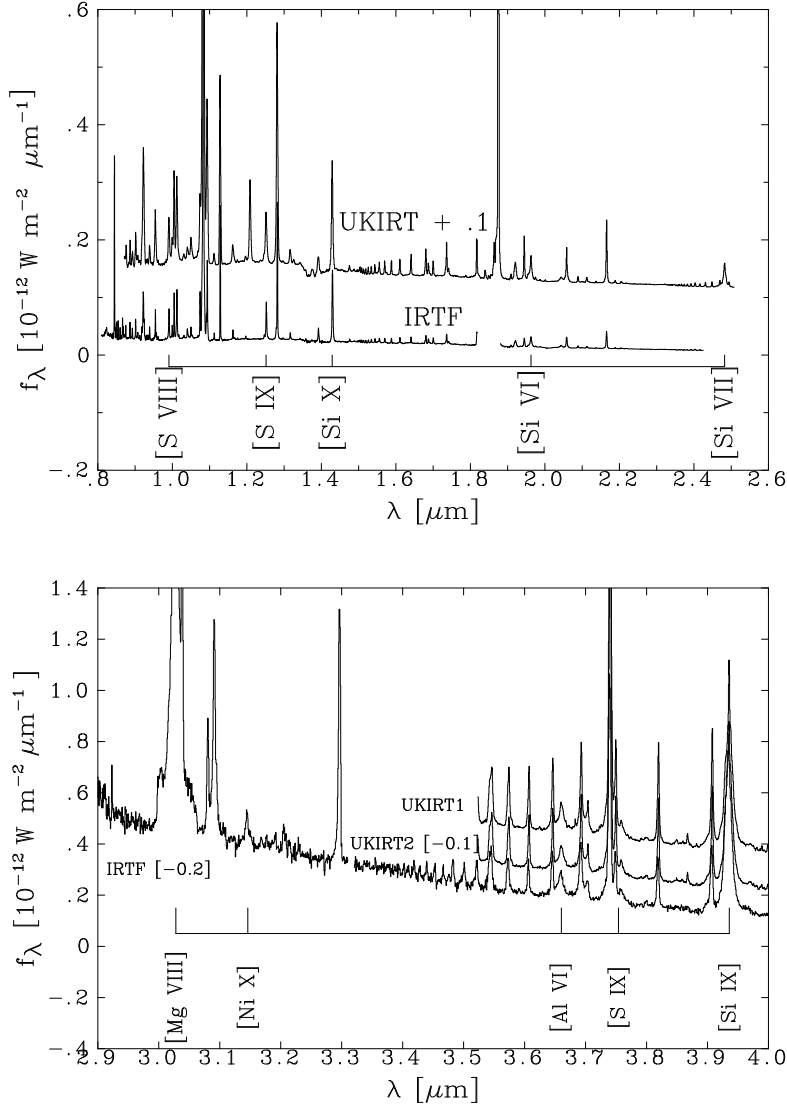

\includegraphics[angle=-90,scale=.40]{ijk.eps}
\includegraphics[angle=-90,scale=.40]{lp_coronal.eps}
\caption{Top: Ground-based (UKIRT and IRTF) observations of \rs, obtained on
2006 April 24 (UKIRT) and May 1 (IRTF); UKIRT spectrum has been displaced upwards
by $0.1\times10^{-12}$~W~m$^{-2}$\mic$^{-1}$. Coronal lines are identified.
Bottom: Ground-based (UKIRT1, UKIRT2 and IRTF) observations of \rs, obtained on
2006 April 16 (UKIRT1), April 24 (UKIRT2) and May 1 (IRTF); spectra UKIRT2 and
IRTF have been displaced downwards by $0.1\times10^{-12}$~W~m$^{-2}$\mic$^{-1}$
and $0.2\times10^{-12}$~W~m$^{-2}$\mic$^{-1}$ respectively. Coronal lines are
identified. Many of the narrow lines without identification are H recombination
lines. All dates UT.}
\label{ijk}
\end{figure}

The SpeX/IRTF data were obtained on 2006 May 1 ($t=77.61$~days) and June 3
($t=110.72$~days), UT using a $0.8''\times15''$ slit and a $10''$ N--S nod for
background cancellation. 
No chopping was performed and extinction corrections were not necessary
because of the proximity of the calibrator star. Data reduction was done
using SpeXTools \citep{cushing} with HD164716 (B9V) as the calibrator star.
The flux of HD164716 was obtained by using the Kurucz \citeyearpar{kurucz1,kurucz2}
model of $\alpha$~Lyr scaled to the $V$ magnitude of HD164716. SpeXTools makes
an automatic, and in this case small, correction for extinction based on the measured
$(B-V)$ colors. UKIRT observations were performed on 2006 April 16 and 24
($t=$~days 62.72 and 70.66 respectively) using the UIST instrument; details
of the observations and data reduction are given in Paper~I. The spectra from both
telescopes, in the 1--2.6\mic\ and 2.8--4.0\mic\ ranges, are shown in
Figs.~\ref{ijk}.

Optical ($BVr'i'z'$) photometry was obtained by the robotic 2m Liverpool 
Telescope \citep[LT;][]{steele}, sited on the island of La Palma, Canaries.
Observations commenced as soon as \rs\ had declined to a level that would not
saturate the detectors; standard photometric reduction techniques were used.

\section{Discussion}

The IR spectrum contains numerous emission lines. In the Spitzer wavelength
range these lines are superimposed on a continuum that declines with increasing
wavelength as $f_\lambda\propto\lambda^{-2}$, consistent with free-free emission.
The emission lines include hydrogen recombination lines, together with fine
structure lines, many of which are coronal;
the hydrogen lines and free-free emission will be discussed elsewhere.
We use flux ratios for lines from the same atomic species and within the same
wavelength band to estimate the electron temperature in the region in which they
originate; the line pairs are listed in Table~\ref{temp}. In Paper~I
the \fion{Si}{vi}/\fion{Si}{x} flux ratio enabled us to estimate the temperature
of the shocked, IR-emitting gas, to be $\simeq9.3\times10^5$~K.

In this paper we have a wide range of ionic species and ionization stages at our
disposal from which to estimate the electron temperature. 
Table~\ref{temp} summarizes the collision strengths ($\Omega$) at
$10^5$~K (Paper~I) for the species and ionization stages observed in the
IR spectra. This temperature is the highest in these sources
available in common to all species and stages of ionization, and the
temperature-dependence of $\Omega$ is relatively weak. The ionization
fractions are from \citet{sutherland}. The \fion{Mg}{vii}5.50\mic\ feature is
blended with the H~16-7 recombination line ($\lambda=5.52$\mic) at the
resolution of the Spitzer SL mode. To correct for this we assume Case~B
\citep{agn}, to estimate the expected flux ratio
$f(\mbox{H16-7})/f(\mbox{H9-7})\simeq0.19$ at electron density $2\times10^5$~cm$^{-3}$
(see below); we use the measured flux in the H~9-7
line ($\lambda=11.31$\mic) to correct the \fion{Mg}{vii} line for the blend
with H~16-7. In view of the uncertainties (particularly in the applicability of
Case~B) we should give the temperature derived from the Mg lines lower weight
than that derived from other species.

The derived temperatures are given in Table~\ref{temp} (column~4). The uncertainties in
$\log{T\mbox{(K)}}$ arising from uncertainties in the line fluxes are $\pm0.2$.
We note that the temperatures in Table~\ref{temp}, derived from the Ne lines
(mean $\simeq1.5\times10^5$~K), are lower than those derived from
Si and Mg lines. The temperature derived from the Si lines in this paper
is not significantly different from that given in Paper~I during the early
($t=55.7$~days) phase. Apparently there is a range of $T$ values present
in the shocked region. This result, derived from analysis of the IR emission lines,
is also suggested by x-ray data obtained on 2006 April 16 \citep{osborne2} and
follows naturally from models of the shock in {\rs} \citep{obrien92}.

We shall assume that the electron
temperature in the region in which the various Ne and \fion{O}{iv} lines
originate is $1.5\times10^5$~K, while that in the Si/Mg line
emitting region is $9\times10^5$~K. The resulting critical densities
$n_{\rm c}$, below which radiative de-excitation begins to dominate
over collisional de-excitation at the assumed temperatures, are given
in Table~\ref{critical}.

We use the fluxes in the \fion{Ne}{v}14.3,24.3\mic\ lines to estimate
the electron density in the `cooler' line-emiting region. Denoting the
$^2$P$_2$, $^2$P$_1$ and $^1$P$_0$ levels as levels `2', `1' and `0'
respectively, the fluxes $f$ in the 14.32\mic\ and 24.3\mic\ lines are 
\begin{equation} f = n_u \: \frac{hc}{\lambda} \: A \: \frac{V}{4\pi D^2} \:\: ,  
\label{Ne}
\end{equation}
where $n_u$ (in cm$^{-3}$, $u=2,1$) is the population of the upper level,
$\lambda$ is the wavelength, $A$ (s$^{-1}$) is the appropriate Einstein
coefficient, $V$ is the emitting volume and $D$ \citep[$=1.6$~kpc;][]{vnu} is
the distance. Assuming that the \fion{Ne}{v}14.2,24.3\mic\ lines arise in the
same region, the flux ratio gives $n_2/n_1\simeq0.6$. 
Detailed balance between radiative de-excitation, and electron collisional
excitation and de-excitation amongst the three levels, together with the above
values for $n_2/n_1$ and the electron temperature, provide a value for the
electron density $n_{\rm e}$. We find $n_{\rm e}\simeq2.2\times10^5$~cm$^{-3}$
in the \fion{Ne}{v} emitting region.

On the basis of radio imaging of the ejecta, \citet{obrien} estimate that the
density in the radio-emitting shell was $\sim10^{-17}$~g~cm$^{-3}$ at
$t\simeq14$~days. The wind/ejecta geometry is known to be complex
\citep{bode07}; however assuming that the density declines as $t^{-2}$
(appropriate for a uniformly expanding shell of constant thickness), we
expect the electron density would be $\sim3\times10^{5}$~cm$^{-3}$ at
$t\simeq63$~days, consistent with our estimate from the Ne lines.
From Table~\ref{critical} we see that this is generally below the critical density
for all Ne lines with the exception of \fion{Ne}{v}24.3\mic, the
weakness of which compared with the other Ne lines may be consistent with a
degree of collisional de-excitation.

We estimate the mass of emitting Ne, assuming that
$n_0+n_1+n_2=n(\mbox{\fion{Ne}{v}})$, and using the \fion{Ne}{v} fluxes from
Table~\ref{critical} and equation~(\ref{Ne}). We find $n(\mbox{\fion{Ne}{v}})
\simeq8.4\times10^{-9}$\Msun\ for a distance of $D=1.6$~kpc and, using the
fractional abundance of \fion{Ne}{v} at $1.5\times10^5$~K from Sutherland \&
Dopita (1993), we get $M(\mbox{Ne})\simeq2\times10^{-7} \,
(D/\mbox{kpc})^2$\Msun. In principle, we could also determine the mass of O;
however it is clear from Table~\ref{critical} that the deduced $n_{\rm e}$ in the
Ne/O region exceeds the $n_{\rm c}$ for the $^2$P$_{3/2}$ state of \fion{O}{iv}.
Consequently we conclude only that $M(\mbox{O}) \gtsimeq 1\times10^{-7} \,
(D/\mbox{kpc})^2$\Msun.

In principle it would be straightforward to use the fluxes in the IR
Ne (and other species) lines, together with the fluxes in the H lines, to
estimate the Ne:H ratio in the ejecta/wind mix. However given the
complex geometry of the ejecta as evidenced by the radio \citep{obrien} and HST
\citep{bode07} observations we do not consider that this is justified until we
have a better understanding of the environment of \rs.
However assuming that the emitting volumes of Ne and O coincide, we find
$n(\mbox{O})/n(\mbox{Ne})\gtsimeq 0.6$, compared with a
$\{n(\mbox{O})/n(\mbox{Ne})\}_\odot$ of 6.6 \citep{asplund}.

\begin{figure}
\includegraphics[angle=-90,scale=.50]{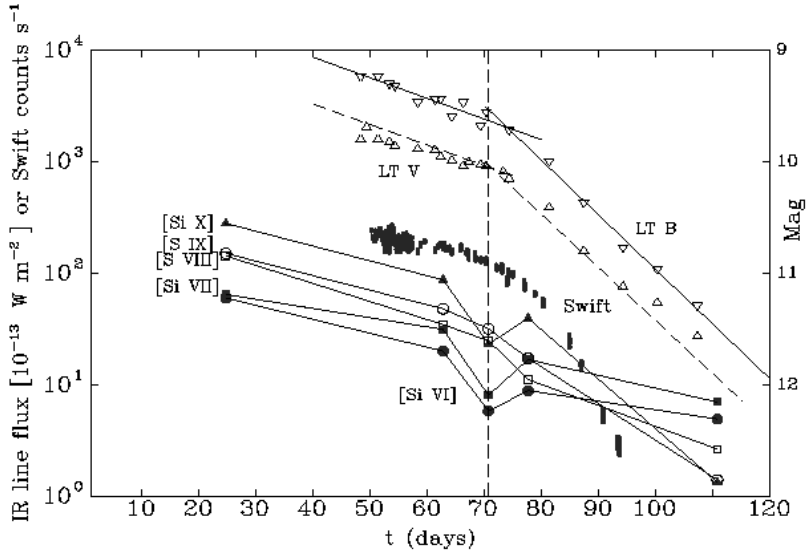}
\caption{Decline in coronal line fluxes ($\bullet=$ \fion{Si}{vi}, $\circ=$
\fion{S}{ix}, $\Box=$ \fion{S}{viii}, $\blacktriangle=$ \fion{Si}{x},
{\small $\blacksquare=$} \fion{S}{viii}); ordinate
(left axis) is line flux in $10^{-13}$~W~m$^{-2}$. Swift fluxes 
in 0.3-10 keV channel (filled points); ordinate (left axis) is in counts
s$^{-1}$. Liverpool Telescope ($BV$) light curves (open triangles; right
axis); $B$ data have been displaced upwards by 1.5~mag for clarity. Straight
lines are least squares fits to light curves for the $40 < t \mbox{~(d)} < 75$,
$75 < t \mbox{~(d)} < 120$ ranges, illustrating break in the light curve around
$t\simeq70$~days. See text for discussion.} 
\label{breakout}
\end{figure}

In Fig.~\ref{breakout} we plot the time-dependence of the coronal line fluxes
for which we have the longest time-base, namely S and Si; line fluxes for
earlier data are taken from Paper~I. It is curious that the S line fluxes
decline monotonically over the first $\sim80$~days of the eruption whereas
there is a distinct minimum in the line fluxes around $t\simeq70$~days
for the Si lines (the errors in the measured fluxes are typically $\simeq5$\%).
We consider that this result is real, because (a)~there is a consistent pattern
in the behavior of the Si and S lines and (b)~it coincides (within the time
resolution) with a distinct `kink' in the $BVr'i'z'$ light curves.
There is no corresponding break in either the x-ray light curve 
\citep[which coincides with the decline of the Super Soft phase;][]{ness}
or in the variation of the hardness ratio \citep[][see Fig.~\ref{breakout},
which shows only the $V$ and $B$ light curves for clarity]{osborne}.

In {\em classical} novae, fluctuations in the visual light curve are often
related to fluctuations in the mass-loss from the WD and the behavior of the
pseudo\-photosphere; however, it is difficult to see how this can also lead to
changes in the coronal line fluxes. Accordingly we consider it unlikely that
this behavior is directly related to the mass-loss.

For the 1985 eruption of \rs, \citet{obrien92} estimated that breakout of
the shock from the edge of the RG wind occurred $\sim60$-70 days after
the eruption, corresponding to 70-80 days after the 2006 eruption, allowing for
the greater inter-eruption interval. While this is close to the day~70
event in the 2006 eruption, it is unlikely that the behavior seen in
Fig.~\ref{breakout} is related to any simple breakout phenomenon, which will be
far more complex than the spherically symmetric model of \citet{obrien92}.
Further, while the electron density we determine is for the Ne/O region, it
is likely that this value is not atypical and the disparity between the
$n_{\rm c}$ (Table~\ref{critical}) and $n_{\rm e}$ values suggest that
the behavior depicted in Fig.~\ref{breakout} is not a collisional
de-excitation effect.
Fig.~\ref{breakout} suggests we may be seeing the combined effects of
recombination, abundance gradients and element segretation, but a detailed
discussion is beyond the scope of this paper.

\section{Concluding remarks}

We have presented the Spitzer IRS spectrum of \rs\ during its 2006 eruption,
the first mid-IR observations of a RN in outburst. The IR spectrum, from
1\mic\ to 30\mic, shows the rich fine structure (including) coronal and
H~recombination line spectrum of the shocked RG wind. There are at least
two temperature regimes in the shocked wind, and the deduced electron
density is consistent with extrapolation from radio observations earlier
in the eruption.

\acknowledgments

This work is based in part on observations made with the Spitzer Space
Telescope, which is operated by the Jet Propulsion Laboratory, California
Institute of Technology under a contract with NASA. We acknowledge the award of
Director's Discretionary Time for this program.
The United Kingdom Infrared Telescope is operated by the Joint Astronomy
Centre on behalf of the U.K. Particle Physics and Astronomy Research Council
(PPARC).
The Liverpool Telescope is operated on the island of La Palma by Liverpool 
John Moores University in the Spanish Observatorio del Roque de los 
Muchachos of the Instituto de Astrofisica de Canarias with financial 
support from the UK Science and Technology Facilities Council.
TRG is supported by the Gemini Observatory, which is operated by the Association
of Universities for Research in Astronomy, Inc., on behalf of the international
Gemini partnership of Argentina, Australia, Brazil, Canada, Chile, the United
Kingdom, and the United States of America.
CEW, AH and RDG are supported by NASA/JPL Spitzer contracts.
The work of DKL, RJR and RWR is supported by The Aerospace
Corporation's Independent Research and Development Program.
MFB was supported by a PPARC Senior Fellowship.
J.-U. N. gratefully acknowledges support provided by NASA through
Chandra Postdoctoral Fellowship grant PF5-60039 awarded by the Chandra
X-ray Center, which is operated by the Smithsonian Astrophysical
Observatory for NASA under contract NAS8-03060.
JPO and KLP acknowledge support from PPARC.
SGS acknowledges partial support from NSF and NASA grants to Arizona State
University.

{\it Facilities:} \facility{Spitzer (IRS)}, \facility{UKIRT}, \facility{IRTF},
\facility{Liverpool Telescope}.

\clearpage

\begin{table}
\begin{center}
\caption{Estimated temperatures from coronal lines.\label{temp}}
\begin{tabular}{lccc}
\tableline\tableline
\multicolumn{1}{c}{Line ratio} & Flux ratio &
                 $\Omega$ at $10^5$~K\tablenotemark{a}
                                    & $\log T$ (K)    \\ \tableline
\fion{Si}{vi}1.96\mic/\fion{Si}{x}1.43\mic & 0.26 & 0.43, 0.13 & 5.96\\
\fion{Mg}{vii}5.50\mic/\fion{Mg}{v}5.61\mic & 3.25 & 1.36, 1.06 & 5.85  \\ 
\fion{Ne}{iii}15.55\mic/\fion{Ne}{v}14.32\mic & 1.18 & 0.79, 2.16  & 5.18\\
\fion{Ne}{iii}15.55\mic/\fion{Ne}{v}24.32\mic & 4.53 & 0.79, 0.75 & 5.15 \\
\fion{Ne}{vi}7.65\mic/\fion{Ne}{iii}15.55\mic & 7.29 & 1.70, 0.79  & 5.34 \\
\fion{Ne}{ii}12.81\mic/\fion{Ne}{v}14.32\mic & 0.99 & 0.40, 2.16 & 5.08  \\
\tableline
\end{tabular}
\tablenotetext{a}{From \citet{SiVI} (\fion{Si}{vi}), 
        \citet{NeIII} (\fion{Mg}{v}, \fion{Ne}{iii}), \citet{NeII} (\fion{Ne}{ii}),
        \citet{NeV} (\fion{Ne}{v}),
        \citet{MgVII} (\fion{Mg}{vii}), \citet{NeVI} (\fion{Ne}{vi}),
         \citet{SiX} (\fion{Si}{x}).}
\end{center}
\end{table}

\begin{table}
\begin{center}
\caption{Critical densities \label{critical}}
\begin{tabular}{lccr}
\tableline\tableline
\multicolumn{1}{c}{Identification} & Term & $\lambda (\!\mic)$  & $n_{\rm c}$ (cm$^{-3}$)
\tablenotemark{a}  
                   \\ \tableline
\fion{Si}{x}    & $^2$P$^o_{3/2}-^2$P$^o_{1/2}$ &   1.4290 & $1.0\times10^{10}$    \\
\fion{Si}{vi}   & $^2$P$^o_{1/2}-^2$P$^o_{3/2}$ &   1.9625 & $1.7\times10^9$   \\
\fion{Si}{vii}  & $^3$P$_0-^3$P$_2$ &    2.4822 & $7.0\times10^8$  \\
\fion{Mg}{viii} & $^2$P$^o_{3/2}-^2$P$^o_{1/2}$ &   3.0269 & $2.6\times10^8$   \\
\fion{S}{ix}    & $^3$P$_0-^3$P$_1$  &   3.7495 & $4.5\times10^8$   \\
\fion{Ne}{ii}   & $^2$P$_{1/2}-^2$P$_{1/3}$ &  12.808  & $2.8\times10^6$  \\
\fion{Ne}{v}    & $^3$P$_2-^3$P$_1$  &  14.311 & $1.4\times10^5$   \\
\fion{Ne}{iii}  & $^3$P$_1-^3$P$_2$ &  15.550   & $8.1\times10^5$ \\
\fion{Ne}{v}    & $^3$P$_1-^3$P$_0$ &  24.300  & $2.4\times10^4$  \\
\fion{O}{iv}    & $^2$P$_{3/2}-^2$P$_{1/2}$  &  25.871  & $2.0\times10^4$   \\
\tableline
\end{tabular}
\tablenotetext{a}{At $1.5\times10^5$~K, for Ne/O lines,
at $9\times10^5$~K for Si/Mg lines; see text.}
\end{center}
\end{table}

\clearpage

\end{document}